 \documentclass[final,1p,times,twocolumn]{elsarticle}

\usepackage{graphics}
\usepackage{graphicx}
\usepackage{epsfig}

\usepackage{amssymb}
\usepackage{amsthm}

\journal{Physica A}

\begin{document}

\begin{frontmatter}



\title{Scaling properties of composite information measures and shape
  complexity for hydrogenic atoms in parallel magnetic and electric fields}

\author[label1]{R Gonz\'alez-F\'erez}
\author[label1]{J S Dehesa}
\author[label2]{S H Patil}
\author[label3]{K D Sen}

\address[label1]{Departamento de F\'{\i}sica At\'omica, Molecular y Nuclear and
  Instituto Carlos I de F\'{i}sica Te\'orica y Computacional, Universidad de
  Granada, 18071-Granada, Spain}
\address[label2]{Department of Physics, Indian Institute of Technology, Mumbai
  400076 India}
\address[label3]{School of Chemistry, University of Hyderabad, Hyderabad 500046
  India}
\ead{sensc@uohyd.ernet.in}


\author{}

\address{}

\begin{abstract}
The scaling properties of various composite information-theoretic
measures (Shannon and R\'enyi entropy sums, Fisher and Onicescu
information products, Tsallis entropy ratio, Fisher-Shannon
product and shape complexity) are studied in position and momentum
spaces for the non-relativistic hydrogenic atoms in the presence
of parallel magnetic and electric fields. Such measures are found
to be invariant at the fixed values of the scaling parameters
given by $ s_1=\frac{B \hbar^3(4\pi\epsilon_0)^2}{Z^2m^2e^3}$ and
$s_2=\frac{F \hbar^4(4\pi\epsilon_0)^3}{Z^3e^5m^2}$. Numerical
results which support the validity of the scaling properties are
shown by choosing the representative example of the position space
shape complexity. Physical significance of the resulting scaling
behaviour is discussed.
\end{abstract}


\end{frontmatter}



\section{Introduction}

The quantum-mechanical uncertainty principle, first formulated
\cite{Heis1} in terms of the standard deviations of the position and momentum
probability densities which characterize the quantum-mechanical
states of one-dimensional single-particle systems, is fundamental
to the understanding the electronic structure and properties of
atoms and molecules. The position-momentum Heisenberg uncertainty
relation has been extensively tested for many three-dimensional
systems \cite{Heis2}, and some interesting properties have been
found for central potentials; namely, the Heisenberg uncertainty
product (i) does not depend on the potential strength for the
bound states of homogeneous power-type potentials
\cite{Katriel01}, and (ii) has a lower bound which has a quadratic
dependence on the orbital quantum number \cite{pablo}. There exist
formulations of the position-momentum uncertainty principle based
on uncertainty measures other than the standard deviation, which
are more stringent than the Heisenberg relation. They are the
uncertainty-like relationships based on, e.g. the Shannon
\cite{Shannon01}, R\'enyi \cite{Renyi01} and Tsallis
\cite{Tsallis01} entropies, the Fisher information
\cite{Fisher01} and the modified LMC or shape complexity, which are found and
discussed in Ref. \cite{Birula01,Birula02,Raja01,jesus,sheila}, respectively.

The scaling properties of the position-momentum uncertainty
relations mentioned above for single particle systems with a wide
variety of central potentials have recently been examined by using
of the dimensional analysis of their associated Schr\"odinger
equation \cite{Katriel01,sen07}. In this letter we present the
first comprehensive information-theoretic study on the
hydrogenic-like atoms in the presence of external parallel
magnetic and electric fields. In particular, we have considered
the scaling properties of the Heisenberg uncertainty measure (i.e.
the standard variation), the Shannon, R\'enyi, Tsallis, Fisher
information measures, and the shape complexity \cite{Lopez95,Lopez02}. The
numerical
validity of these scaling properties are presented. The predictive
power of the presently obtained results on this statistical
complexity is illustrated by taking the example of the most
distinctive non-linear spectroscopic phenomenon, the avoided
crossing of two energy levels with the same energy \cite{wigner}
of a hydrogenic system in the presence of intense parallel
magnetic and electric fields.

This paper is organized as follows. We analyze the
scaling transformation of the energies and eigenfunctions which
characterize the quantum-mechanical states of a hydrogenic atom in
the presence of parallel magnetic and electric fields in Section
2, and the dimensional properties of their position and momentum
Heisenberg uncertainty measure in Section 3. In Section 4, we
examine the scaling properties of the uncertainty relations
associated with the following information-theoretic measures:
Shannon, R\'enyi and Tsallis entropies and the Fisher and Onicescu
informations as well as the shape complexity. Finally, in Section
5, we compute the shape complexity for two different pairs of
energy levels of a hydrogen atom under intense parallel magnetic
and electric fields, which show avoided crossing phenomena.
Moreover, we check the validity of the corresponding scaling law
obtained for this information measure in the previous section,
and, most important, we show that this measure presents a peculiar
mirror symmetry through the avoided crossing region. The latter
implies that the shape complexity is a good indicator of this
highly non-linear phenomena at the same level as the energy
\cite{wigner,ruder,peter98} and the Shannon and Fisher
informations \cite{gonzalez2003}.

\section{Hydrogenic systems in parallel magnetic and electric fields: scaling
properties}

Let us consider an electron moving in a Coulombic potential due to a
nucleus with charge $+Ze$, in the presence of parallel magnetic and
electric fields oriented in the $z$ direction. The effective potential in
spherical coordinates is
\begin{equation}\label{2.1}
V(\mathbf{r})=-\frac{Ze^2}{4\pi\epsilon_0 r}+\frac{eB}{m}L_z
+\frac{e^2B^2}{2m}r^2\sin^2\theta+eFr\cos\theta.
\end{equation}
where $B$ and $F$ are the constant magnetic and electric fields strengths, $m$
the mass of the electron, $\epsilon_0$ the electric constant, and
$L_z$ the $z$-component of the angular momentum. The
corresponding Schr\"{o}dinger equation for this potential is
\begin{equation}\label{2.2}
-\frac{\hbar^2}{2m}\nabla^2 \psi+V(\mathbf{r})\psi=E\psi.
\end{equation}
Note that we have neglected the relativistic corrections \cite{chen}
and the spin-orbit coupling \cite{garstang} and assumed that the nuclear mass
is infinity \cite{palovzhilinski}, because they do not affect the main results
of this work.
Here, we shall study the scaling properties of the eigensolutions of this
equation of motion under the transformation $\mathbf{r}= \lambda\mathbf{r}'$,
where the scaling factor is specified later on. Then the transformed
Schr\"odinger equation takes on the form
\begin{displaymath}
-\frac{\hbar^2}{2m\lambda^2}\nabla'^2 \psi+
\left[-\frac{Ze^2}{4\pi\epsilon_0 \lambda r'}+\frac{eB}{m}L_{z'}
+\frac{e^2B^2 \lambda^2}{2m}r'^2\sin^2\theta'+eF \lambda r'\cos\theta'
\right]
\psi=E\psi. \nonumber
\end{displaymath}
Now, we multiply this equation by the factor
$\frac{m\lambda^2}{\hbar^2}$ and fix $\lambda$ by imposing that
the factor in the Coulomb term is equal to unity. It turns out
that
\begin{equation}\label{lambda}
\lambda= \frac{4\pi\epsilon_0\hbar^2}{Ze^2 m},
\end{equation}
and the scaled Schr\"odinger equation reads as
\begin{displaymath}
\left[-\frac{1}{2}\nabla'^2-\frac{1}{r'}+\frac{s_1}{\hbar}L_{z'}
+\frac{s_1^2 }{2}r'^2\sin^2\theta'+s_2 r'\cos\theta'\right]\psi =E_1\psi,
\end{displaymath}
where
\begin{equation}
  \label{eq:s1_def}
   s_1=\frac{B \hbar^3(4\pi\epsilon_0)^2}{Z^2m^2e^3}, \quad \rm{and}
\quad s_2=\frac{F \hbar^4(4\pi\epsilon_0)^3}{Z^3e^5m^2}.
\end{equation}
Note that $\lambda$ has length units and the new coordinate is dimensionless,
as wanted. Moreover, the parameter $s_1$ and $s_2$ are also dimensionless, and
the energy $E(\hbar^2/m,Z,B,F)$ rescales into $E_1=E(1,1,s_1,s_2)$ as
\begin{displaymath}
E(\hbar^2/m,Z,B,F)=\frac{e^4Z^2m}{\hbar^2(4\pi\epsilon_0)^2}E(1,1,s_1,s_2).
\end{displaymath}
Consequently the wavefunction $\psi(\mathbf{r};\hbar^2/m,Z,B,F)$ will change as
\begin{displaymath}
  \label{eq:wf}
\psi(\mathbf{r};\hbar^2/m,Z,B,F)= \lambda^{-3/2}\psi(\mathbf{r}';1,1,s_1,s_2),
\end{displaymath}
because of the normalization to unity, and the associated probability density
$\rho(\mathbf{r})=|\psi(\mathbf{r})|^2$ as
\begin{equation}
  \label{eq:pd}
\rho(\mathbf{r};\hbar^2/m,Z,B,F)= \lambda^{-3}\rho(\mathbf{r}';1,1,s_1,s_2).
\end{equation}

To obtain the scaling of the wavefunction in momentum space,
$\tilde\psi(\mathbf{p}\def\tilde\psi(\mathbf{p};\hbar^2/m,Z,B,F)$, under the
transformation $\mathbf{p}'= \lambda\mathbf{p}$, with $\lambda$ given by
(\ref{lambda}), we take into account that $\psi(\mathbf{r})$ and
$\tilde\psi(\mathbf{p})$ are mutually Fourier-transformed as
\begin{displaymath}\label{eq:fouriertrans}
\tilde\psi(\mathbf{p};\hbar^2/m,Z,B,F)=\frac{1}{(2\pi\hbar)^{3/2}}\int
e^{-i\mathbf{p}\cdot \mathbf{r}/\hbar}\psi(\mathbf{r};\hbar^2/m,Z,B,F)
d\mathbf{r}.
\end{displaymath}
It is straightforward that the momentum wavefunction scales as
\begin{displaymath}
  \label{eq:wf_ms}
\tilde\psi(\mathbf{p};\hbar^2/m,Z,B,F)=\lambda^{3/2}
\tilde\psi(\mathbf{p}';1,1,s_1,s_2)
\end{displaymath}
and the associated density $\gamma(\mathbf{p})=|\tilde\psi(\mathbf{p})|^2$ as
\begin{equation}
  \label{eq:pd_ms}
\gamma(\mathbf{p};\hbar^2/m,Z,B,F)= \lambda^{3}\gamma(\mathbf{p}';1,1,s_1,s_2).
\end{equation}

\section{Dimensionality properties of the Heisenberg uncertainty measure}
For a hydrogenic system with a potential $V(\mathbf{r})$ given by
Eq. (\ref{2.1}), a pure dimensional analysis of the standard deviation of its
position wavefunction $\psi(\mathbf{r})$ defined by
\begin{displaymath}
\label{eq:deviation}
\sigma_{\mathbf{r}}^2=\int \psi^*(\mathbf{r})
(\mathbf{r}-\langle\mathbf{r}\rangle)^2\psi(\mathbf{r})d\mathbf{r},
\end{displaymath}
allows us to write down in a straightforward maner that
\begin{equation}\label{2.5}
\sigma_{\mathbf{r}}=\lambda  f_1(s_1,s_2),\qquad
\end{equation}
where $f_1(s_1,s_2)$ is a fixed function of the dimensionless parameters $s_1$
and $s_2$ given by Eq. (\ref{eq:s1_def}). Moreover, taking into account the
reciprocity of the position and momentum spaces, a similar dimensional
analysis for the standard deviation in momentum space
\begin{displaymath}
  \label{eq:deviation_p}
  \sigma_{\mathbf{p}}^2=\int \tilde\psi^*(\mathbf{p})
(\mathbf{p}-\langle\mathbf{p}\rangle)^2\tilde\psi(\mathbf{p})d\mathbf{p},
\end{displaymath}
leads to the expression
\begin{equation}\label{2.5.2}
\sigma_{\mathbf{p}}=\hbar \lambda^{-1}f_2(s_1,s_2).
\end{equation}
Hence, the Heisenberg uncertainty product is
\begin{equation}\label{2.6}
\sigma_{\mathbf{r}}\;\sigma_{\mathbf{p}}=\hbar~f_1(s_1) f_2(s_2).
\end{equation}
Expressions (\ref{2.5}), (\ref{2.5.2}) and (\ref{2.6}) allow us to state for
hydrogenic systems under parallel magnetic and electric fields, that (i) the
position and momentum spreadings around the corresponding centroids in
position and momentum space depend only on the nuclear charge $Z$ and the
dimensionless parameters $s_1$ and $s_2$, and (ii) the Heisenberg uncertainty
product depends only on $s_1$ and $s_2$.

\section{Scaling of hydrogenic information-theoretic uncertainty measures}
Here we examine the scaling properties of the information-theoretic-based
uncertainty measures of Shannon, Fisher, Onicescu and Tsallis types, as well
as their mutual relations, under the coordinate transformation
$\mathbf{r}=\lambda\mathbf{r}'$ (where the scaling $\lambda$ is given in
Eq. (\ref{lambda})) for a hydrogenic system in the presence of parallel
magnetic and electric fields. In particular, we show that the Shannon entropy
sum, the Fisher and Onicescu information products, the Tsallis entropy ratio,
the Fisher-Shannon measure and the shape complexity of this system
depend only on the dimensionless parameters $s_1$ and $s_2$ for given values
of the nuclear charge $Z$ and the strengths
$(B,F)$ of the external fields as described by Eqs. (\ref{3.15}),
(\ref{3.18}), (\ref{3.24}), (\ref{3.21}), (\ref{3.28}),
(\ref{3.32})-(\ref{3.33}), and (\ref{3.34})-(\ref{3.35}),
respectively, later on.

\subsection{Shannon entropy sum}
The Shannon entropies \cite{Shannon01} in the position space and momentum
space, are
\begin{displaymath}\label{3.13}
S_r=-\int \rho(\mathbf{r})\ln\rho(\mathbf{r})d\mathbf{r}, \qquad
S_p=-\int \gamma(\mathbf{p})\ln\gamma(\mathbf{p})d\mathbf{p}.
\end{displaymath}
Using the relations in Eqs. (\ref{eq:pd}) and (\ref{eq:pd_ms}), we get for
these entropies the scaling properties
\begin{eqnarray}\label{3.14}
&&S_r(\hbar^2/m,Z,B,F)=3\ln\lambda+S_r(1,1,s_1,s_2),\\
&&S_p(\hbar^2/m,Z,B,F)=-3\ln\lambda+S_p(1,1,s_1,s_2),
\nonumber
\end{eqnarray}
which imply that the Shannon entropy sum $S_T=S_r+S_p$ satisfies
the relation
\begin{equation}\label{3.15}
S_T(\hbar^2/m,Z,B,F)=S_T(1,1,s_1,s_2).
\end{equation}

\subsection{Fisher information product}
The Fisher information \cite{Fisher01} measures for position and momentum are
\begin{displaymath}\label{3.16}
I_r=\int \frac{[\mathbf{\nabla} \rho(\mathbf{r})]^2}{\rho(\mathbf{r})}
d\mathbf{r}, \qquad
I_p=\int \frac{[\mathbf{\nabla} \gamma(\mathbf{p})]^2}{\gamma(\mathbf{p})}
d\mathbf{p}.
\end{displaymath}
Using the relations in Eqs (\ref{eq:pd}) and (\ref{eq:pd_ms}), one obtains
the scaling properties
\begin{eqnarray}\label{3.17}
&&I_r(\hbar^2/m,Z,B,F)=\frac{1}{\lambda^2}I_r(1,1,s_1,s_2),\\
&&I_p(\hbar^2/m,Z,B,F)=\lambda^2I_p(1,1,s_1,s_2),\nonumber
\end{eqnarray}
which together imply that the Fisher information product
$I_{rp}=I_rI_p$ satisfies the relation
\begin{equation}\label{3.18}
I_{rp}(\hbar^2/m,Z,B,F)=I_{rp}(1,1,s_1,s_2).
\end{equation}

\subsection{Onicescu information product}
The Onicescu informations \cite{Onicescu01} in position and momentum spaces are
\begin{displaymath}\label{3.22}
E_r=\int [\rho(\mathbf{r})]^2 d\mathbf{r},\qquad E_p=
\int [\gamma(\mathbf{p})]^2d\mathbf{p}.
\end{displaymath}
Using the relations in Eqs. (\ref{eq:pd}) and (\ref{eq:pd_ms}), we get
the scaling properties
\begin{eqnarray}\label{3.23}
&&E_r(\hbar^2/m,Z,B,F)=\frac{1}{\lambda^3}E_r(1,1,s_1,s_2),\\
&&E_p(\hbar^2/m,Z,B,F)=\lambda^3E_p(1,1,s_1,s_2),\nonumber
\end{eqnarray}
which imply that the Onicescu information product $E_{rp}=E_rE_p$
satisfies the relation
\begin{equation}\label{3.24}
E_{rp}(\hbar^2/m,Z,B,F)=E_{rp}(1,1,s_1,s_2).
\end{equation}

\subsection{R\'{e}nyi entropy sum}
The R\'{e}nyi entropies \cite{Renyi01} in position and momentum spaces are
\begin{displaymath}\label{3.19}
H_{\alpha}^{(r)}=\frac{1}{1-\alpha}\ln \int
[\rho(\mathbf{r})]^{\alpha} d\mathbf{r}, \qquad H_{\alpha}^{(p)}=\frac{1}
{1-\alpha}\ln \int [\gamma(\mathbf{p})]^{\alpha} d\mathbf{p}.
\end{displaymath}
With the relations in Eqs. (\ref{eq:pd}) and (\ref{eq:pd_ms}), we get for
these entropies the scaling properties
\begin{eqnarray*}\label{3.20}
&&H_{\alpha}^{(r)}(\hbar^2/m,Z,B,F)=3\ln \lambda
+H_{\alpha}^{(r)}(1,1,s_1,s_2), \\
&&H_{\alpha}^{(p)}(\hbar^2/m,Z,B,F)=-3\ln \lambda
+H_{\alpha}^{(p)}(1,1,s_1,s_2), \nonumber
\end{eqnarray*}
which imply that the R\'{e}nyi entropy sum $H_{\alpha}^{(T)}
=H_{\alpha}^{(r)}+H_{\alpha}^{(p)}$ satisfies the relation
\begin{equation}\label{3.21}
H_{\alpha}^{(T)}(\hbar^2/m,Z,B,F)=H_{\alpha}^{(T)}(1,1,s_1,s_2).
\end{equation}

\subsection{Tsallis entropy ratio}
The Tsallis entropies \cite{Tsallis01} in position and momentum spaces are
\begin{displaymath}\label{3.25}
T^{(r)}_n=\frac{1}{n-1}\Big[1-J^{(r)}_n\Big],\quad
T^{(p)}_q=\frac{1}{q-1}\Big[1-J^{(p)}_q\Big], \quad
\frac{1}{q}+\frac{1}{n}=2.
\end{displaymath}
where the integral terms are given by 
\begin{displaymath}\label{3.26}
J^{(r)}_n
=\int[\rho(\mathbf{r})]^n d\mathbf{r}, \quad
J^{(p)}_q
=\int[\gamma(\mathbf{p})]^q d\mathbf{p}.
\end{displaymath}
Using the relations in Eqs. (\ref{eq:pd}) and (\ref{eq:pd_ms}), we get
the scaling properties
\begin{eqnarray}\label{3.27}
&&J^{(r)}_n(\hbar^2/m,Z,B,F)=\lambda^{3-3n}J^{(r)}_n(1,1,s_1,s_2), \nonumber \\
&&J^{(p)}_q(\hbar^2/m,Z,B,F)=\lambda^{3q-3}J^{(p)}_q(1,1,s_1,s_2).  \nonumber
\nonumber
\end{eqnarray}
Then one obtains for the ratio
$J_{p/r}=\frac{(J^{(p)}_q)^{1/2q}}{(J^{(r)}_n)^{1/2n}}$
the following equality
\begin{eqnarray}\label{3.28}
&&J_{p/r}(\hbar^2/m,Z,B,F)=J_{p/r}(1,1,s_1,s_2), \quad
\frac{1}{n}+\frac{1}{q}=2.
\end{eqnarray}

\subsection{Fisher-Shannon measure}
For the Shannon entropy power
\begin{displaymath}\label{3.29}
N_r=\frac{1}{\pi e}e^{2S_r/3},\qquad N_p=\frac{1}{\pi e}e^{2S_p/3}
\end{displaymath}
in the two conjugated spaces,
we obtain from Eqs. (\ref{3.14}) the following scaling:
 \begin{displaymath}\label{3.30}
 N_r(\hbar^2/m,Z,B,F)= \lambda^2 N_r(1,1,s_1,s_2),
 \end{displaymath}
\begin{displaymath}\label{3.31}
N_p(\hbar^2/m,Z,B,F)=\frac{1}{\lambda^2}N_p(1,1,s_1,s_2),
\end{displaymath}
Using these expressions and  Eq. (\ref{3.17}) for the Fisher information, we obtain for the
Fisher-Shannon measure the scaling
\begin{equation}\label{3.32}
N_r(\hbar^2/m,Z,B,F) I_r(\hbar^2/m,Z,B,F)=
N_r(1,1,s_1,s_2)I_r(1,1,s_1,s_2),
\end{equation}
\begin{equation}\label{3.33}
N_p(\hbar^2/m,Z,B,F) I_p(\hbar^2/m,Z,B,F)=
N_p(1,1,s_1,s_2)I_p(1,1,s_1,s_2)
\end{equation}
in position and momentum spaces, respectively.

\subsection{Shape complexity}
For the shape complexity  \cite{Lopez95,Lopez02} $C=e^{S_r} E_r$, with
$S_r$  and $E_r$ being the Shannon entropy and the Onicescu information or disequilibrium,
respectively, we use the relations in Eqs. (\ref{3.14}) and (\ref{3.23}) to
obtain
\begin{equation}\label{3.34}
e^{S_r(\hbar^2/m,Z,B,F)} E_r(\hbar^2/m,Z,B,F)
=e^{S_r(1,1,s_1,s_2)} E_r(1,1,s_1,s_2),
\end{equation}
\begin{equation}\label{3.35}
e^{S_p(\hbar^2/m,Z,B,F)} E_p(\hbar^2/m,Z,B,F)
=e^{S_p(1,1,s_1,s_2)} E_p(1,1,s_1,s_2),
\end{equation}
for the scaling in the two reciprocal spaces.

Besides the scaling invariance shown by Eqs.  (\ref{3.34}) and
(\ref{3.35}) for the shape complexity, there are two noteworthy
features: (i) the scaling properties are independent of the
relative orientation of the external fields, and more
interestingly, (ii) the functional dependence on $s_1$ and $s_2$
predicts the existence of extremum points when one of the fields
is varied keeping  fixed the other one. We note here that the
functional form of the shape complexity is not obtained through
the dimensional analysis and the number of maximum and minimum
points in it depends upon the specific details. In the next
section we shall discuss these features in some detail. These
observations are equally valid for the other uncertainty-like
products discussed in this work.

\section{Hydrogenic shape complexity: numerical scaling test and avoided
crossing indicator} We have successfully carried out extensive
numerical tests of the scaling properties of the various
uncertainty-like products discussed above. In this section, we
will use atomic units ($m=\hbar=e=4\pi\epsilon_0=1$) and take $B$
in units of speed of light $c$. We will discuss the shape
complexity, as a representative example, in the neighborhood of
some typical avoided crossings of hydrogenic systems in parallel
magnetic and electric field. The details of the computational
approach used to solve the Schr\"odinger equation (\ref{2.2}) can
be found elsewhere \cite{jcam}. In particular, we have considered
the pair of levels $3p_0$ and $3d_0$ of the ($Z=1$) hydrogen atom,
for which the paramagnetic term does not contribute. Note that,
for simplicity, the field-free quantum numbers are used to label
these states. In the presence of the magnetic field the magnetic
quantum number and the $z$-axis parity are good quantum numbers.
Hence, these levels have different symmetry and as the magnetic
field strength is varied they could have the same energy, which
occurs at the magnetic field interval $0.087$ a.u. $\le B \le
0.08825$ a.u.

\begin{figure}
\includegraphics[scale=0.75]{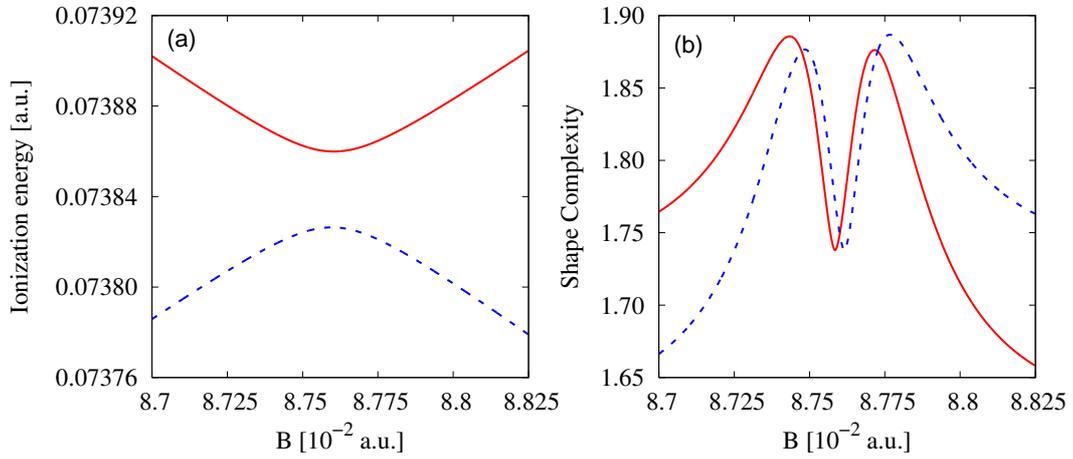}
\caption{\label{fig:ac_3p3d_z_1} Color online. The ionization
energies (a) and shape complexities (b) of the states $3p_0$
(dashed line) and $3d_0$ (solid line),  of the hydrogen atom
(so, with $Z=1$) in parallel electric and magnetic fields as a function
of the magnetic field strength, and with an electric field fixed to
$F=1.946 \times 10^{-6}$ a.u.}
\end{figure}

\begin{figure}
\includegraphics[scale=0.75]{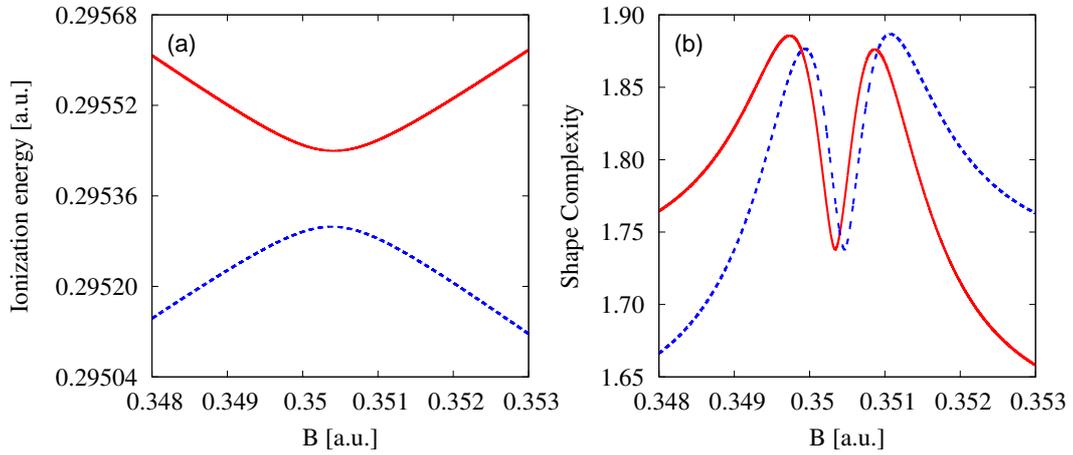}
\caption{\label{fig:ac_3p3d_z_2}
Color online. The same as Fig. \ref{fig:ac_3p3d_z_1}, but for a hydrogenic
atom with nuclear charge $Z=2$ and $F=1.557\times 10^{-5}$ a.u. }
\end{figure}

If an additional parallel electric field is also on, only the azimuthal
symmetry remains so that both levels may have the same symmetry; then an
avoided crossing is formed between them due to the Wigner-non-crossing rule
\cite{wigner}. This non-linear phenomenon is illustrated in
Figs. \ref{fig:ac_3p3d_z_1}a and b, which show the ionization energies
and shape complexities, respectively, of these levels for a magnetic field
with strength $0.087$ a.u.$ \le B_1 \le 0.08825$ a.u. and a fixed electric
field with strength $F_1=1.946 \times 10^{-6}$ a.u. An analogous result
should be expected for the same pair of states in a hydrogenic atom with
nuclear charge $Z=2$ if the magnetic and electric field strengths are
scaled according the rules discussed in the previous section. The
corresponding energies and shape complexities are presented in
Fig. \ref{fig:ac_3p3d_z_2}a and b, as a function of the magnetic
field strength in the range $0.348$ a.u. $\le B_2\le 0.353$ a.u., and
fixed electric field strength $F_2=1.557\times 10^{-5}$ a.u. Note, that the
scaling laws $ F_2=F_1*(Z=2)^3$ and $B_2=B_1*(Z=2)^2$ are satisfied.

Let us first analyze the ionization energy. Looking at Figs.
\ref{fig:ac_3p3d_z_1}a and \ref{fig:ac_3p3d_z_2}a, the ionization
energy shows a qualitatively similar but quantitatively different
behavior as a function of $B$ in the two hydrogenic atoms. On the
one hand, the typical avoided-crossing behavior is observed, i.e.
they approach each other with increasing magnetic field, until
they come close and strongly interact, splitting apart thereafter.
For both systems, the ionization energy of the $3p_0$ ($3d_0$)
state monotonically increases (decreases) as the magnetic field
strength is enhanced, passes through a maximum (minimum), and
decreases (increases) thereafter. However, major differences
appear in the computed values of the energies, which differ by a
factor $Z^2$, as expected by the scaling properties discussed
above. The minimal energetic spacing $\Delta
E=|E_{3p_0}-E_{3d_0}|=3.35 \times 10^{-5}$ a.u. occurs at the
field strength $B= 8.760038 \times 10^{-2}$ a.u. for the $Z=1$
atom. For the $Z=2$ atom, the avoided crossing is energetically
much broader, being $\Delta E=|E_{3p_0}-E_{3d_0}|=1.4 \times
10^{-4}$ a.u. the minimal energetic spacing at $B=0.3504$ a.u.
Please note the different energy scales in Figs.
\ref{fig:ac_3p3d_z_1}a and  \ref{fig:ac_3p3d_z_2}a.

The evolution of the shape complexities with the magnetic field,
as can be seen from Figs. \ref{fig:ac_3p3d_z_1}b and
\ref{fig:ac_3p3d_z_2}b, displays interesting features. They show a
double-hump structure with a mirror symmetry as a function of the
magnetic field strength. The computed values for the shape
complexity are identical, although they are achieved at the
different magnetic field strengths which are related by the
scaling rules as derived above. Close to the magnetic field
strength at which the minimal energetic spacing occurs, the shape
complexities of both states achieve the same value,
$C_{3p_0}=C_{3d_0}=1.7492$, and this is at $B=0.0876004$ a.u. and
$0.350416$ a.u. for the $Z=1$ and $2$ systems, respectively. The
minimal values of the shape complexities are equal for both
states, $C_{3p_0}=C_{3d_0}=1.7380$, and are located at symmetric
positions with respect to the critical magnetic field values
$B_c$, i.e. the $3p_0$ and $3d_0$ minima are shifted to the left
and to the right by $1.569 \times 10^{-5}$ a.u. and $1.566\times
10^{-5}$ a.u. for the $Z=1$ system, and by $6.3\times 10^{-5}$
a.u. and $6.2\times 10^{-5}$ a.u. for the $Z=2$ atom,
respectively. The first hump of $C_{3d_0}$ and the second one of
$C_{3p_0}$, also have the very similar value $C_{3d_0}=1.8856$ and
$C_{3p_0}=1.8867$, and are shifted to the left and to the right by
$1.6706 \times 10^{-4}$  a.u. and $1.6632 \times 10^{-4}$ a.u. for
the $Z=1$ atom, respectively, and by $6.64 \times 10^{-4}$ a.u.
and $6.69 \times 10^{-4}$ a.u. for the $Z=2$ system, respectively.
Analogously, the second maxima of the $3p_0$ level,
$C_{3p_0}=1.8766$, and first one of the  $3d_0$ state,
$C_{3d_0}=1.8762$, are identical for both systems, and are shifted
for $Z=1$ by $1.1591 \times 10^{-4}$  a.u. and $1.1566 \times
10^{-4}$ a.u., to the right and left, respectively; and for $Z=2$
are shifted $4.64 \times 10^{-4}$ a.u. and $4.62 \times 10^{-4}$
a.u. to the right, respectively.
It is interesting to remark that the presently calculated values of the shape
complexity $C$ obey the universal bound $C\ge 1$, which has been recently
shown for general monodimensional
\cite{Lopez95} and $D$-dimensional ($D\ge1$) probability
densities \cite{sheila}.

Finally, let us point out here
that in absence of the external fields, for the free
hydrogenic atoms, $C$ is a constant, independent of the nuclear
charge $Z$. This is a consequence of the homogeneous character of
the potential which leads to a parameter-free scaling property of
the shape complexity \cite{sen07}. In presence of the external
fields, the shape complexity varies with the parameters of the
potential which becomes inhomogeneous in character.

In conclusion, the existence of extremum points and the scaling behavior with
the external fields is numerically verified for the shape complexity
as given in Eq. (\ref{3.34}). Further, according to the shape
complexity analysis here described the scaling property can be used to predict
the existence of avoided crossings for a heavy hydrogenic atom under strong
external fields from the avoided crossings data on a lighter member and
vice-versa. Similar results should be expected for the remaining composite
information-measures analyzed in this work.

\section*{Acknowledgements}
Financial support by the Spanish
projects FIS2008--02380 (MICINN) and grants
P06--FQM--01735 and FQM-2445 (Junta de Andaluc\'{\i}a) is
gratefully appreciated. S.H.P. acknowledges support from
A.I.C.T.E. as emeritus fellow. RGF and JSD belong to the Andalusian research
group FQM-207.





\end{document}